\documentclass[journal,twoside]{IEEEtran}
\usepackage[normalem]{ulem}
\usepackage{cite}

\ifCLASSINFOpdf
   \usepackage[pdftex]{graphicx}
\else
\fi
\usepackage{amsmath,amsfonts}
\usepackage{algorithm}
\usepackage{enumitem}
\usepackage{algorithmic}
\usepackage{caption}

\usepackage{array}
\usepackage{tabularx}
\usepackage{hyperref}
\usepackage{amssymb}
\usepackage{cleveref}
\usepackage{tikz}
\usepackage{multirow}
\usepackage{mathrsfs}
\usepackage{upgreek}
\usepackage{afterpage}
\usepackage{ulem}

\usepackage{float}

\newcolumntype{M}{>{\centering\arraybackslash}m{\dimexpr.25\linewidth-2\tabcolsep}}

\usetikzlibrary{shapes,arrows,positioning}
\definecolor{dark_blue}{RGB}{46,87,144}
\definecolor{blue_pers}{RGB}{54,104,171}
\definecolor{grey_pers}{RGB}{245,245,245}
\definecolor{red_pers}{RGB}{213,78,33}
\definecolor{green_pers}{RGB}{194, 239, 194}

\definecolor{ppink}{RGB}{236, 0, 84}
\definecolor{porange}{RGB}{254, 64, 18}
\definecolor{pgreen}{RGB}{0, 0, 70}
\definecolor{pblue}{RGB}{44, 95, 177}

\usepackage{xcolor}

\definecolor{dark_blue}{RGB}{46,87,144}
\definecolor{dark_green}{RGB}{0,100,0}

\newcommand\revn[1]{\textcolor{dark_green}{\sout{#1}}}

\tikzstyle{block} = [rectangle, draw, fill=grey_pers!30, 
    text width=6.5em, text centered, rounded corners, minimum height=4em]

\tikzstyle{type} = [rectangle, draw, fill=grey_pers, 
    text width=6.5em, text centered, minimum height=4em]    
\tikzstyle{objection} = [rectangle, draw, 
    text width=6.5em, text centered, rounded corners, minimum height=4em]
\tikzstyle{line} = [draw, -latex']
\tikzstyle{line_1} = [draw]

\tikzstyle{cloud} = [rectangle, draw, fill = pgreen, draw=none,  rounded corners,
    text width=5em, text centered, minimum height=1.5em]

\tikzstyle{circle_blue} = [draw,circle,fill=grey_pers!30, node distance=2cm,
    minimum size=2em]
    \tikzstyle{circle_green} = [draw,circle,
     node distance=2cm,
    minimum size=2em]
\tikzstyle{circle_grey} = [draw,circle,fill=grey_pers, node distance=2cm,
    minimum size=2.2em] 
\tikzstyle{circle_red} = [draw,circle,fill=red_pers!30, node distance=2cm,
    minimum size=2em] 
\tikzstyle{back} = [rectangle, draw, 
    text width=7em, text centered, minimum height=1.5em]

\tikzstyle{back_b} = [rectangle, draw, fill=pblue, draw=none, rounded corners,
    text width=5em, text centered, minimum height=1.5em]

\tikzstyle{back_g} = [rectangle, draw=none, fill=pgreen,  rounded corners,
    text width=5em, text centered,  minimum height=1.5em]

\newcommand{\be}{\begin{equation}}
\newcommand{\ee}{\end{equation}}

\newcommand{\bmI}{\boldsymbol{\mathcal{I}}}
\newcommand{\bmR}{\boldsymbol{\mathcal{R}}}

\newcommand{\bmA}{\boldsymbol{\mathcal{A}}}
\newcommand{\hbmA}{ \boldsymbol{ \hat{\mathcal{A}}}}

\newcommand{\hbmS}{\boldsymbol{\hat{\mathcal{S}}}}

\newcommand{\bE}{{\bf E}}
\newcommand{\bF}{{\bf F}}

\newcommand{\bL}{\boldsymbol{L}}

\newcommand{\bH}{\boldsymbol{H}}

\newcommand{\bU}{{\bf U}}

\newcommand{\bZ}{{\bf Z}}

\newcommand{\bee}{{\bf e}}
\newcommand{\bg}{{\bf g}}
\newcommand{\bb}{{\bf b}}
\newcommand{\bxi}{\boldsymbol{\xi}}

\newcommand{\bphi}{\boldsymbol{\phi}}

\newcommand{\bd}{{\bf d}}
\newcommand{\bq}{{\bf q}}
\newcommand{\bp}{{\bf p}}
\newcommand{\bj}{{\bf j}}
\newcommand{\bm}{{\bf m}}

\newcommand{\bn}{{\bf n}}

\newcommand{\bh}{{\bf h}}
\newcommand{\bu}{{\bf u}}
\newcommand{\bv}{{\bf v}}

\newcommand{\bff}{{\bf f}}
\newcommand{\bx}{{\bf x}}
\newcommand{\by}{{\bf y}}

\newcommand{\fD}{\mathsf{D}}
\newcommand{\fK}{\mathsf{K}}

\newcommand{\mT}{\mathcal{T}}
\newcommand{\mA}{\mathcal{A}}
\newcommand{\mH}{\mathcal{H}}

\newcommand{\mK}{\mathcal{K}}
\newcommand{\mI}{\mathcal{I}}

\newcommand{\mO}{\mathcal{O}}
\newcommand{\mM}{\mathcal{M}}

\newcommand{\mU}{\mathcal{U}}

\newcommand{\mE}{\mathcal{E}}
\newcommand{\mR}{\mathcal{R}}

\newcommand{\IV}{\mathbb{V}}

\newcommand{\IE}{\mathbb{E}}
\newcommand{\IP}{\mathbb{P}}
\newcommand{\IR}{\mathbb{R}}

\newcommand{\bcurl}{\operatorname{\bf{curl}}}

\newcommand{\divg}{\operatorname{div}}

\newcommand{\bgamma}{{\boldsymbol{\gamma}}}
\newcommand{\bSigma}{{\boldsymbol{\Sigma}}}

\newcommand{\e}{\operatorname{e}}

\definecolor{myb}{RGB}{47,85,161}
\definecolor{myg}{RGB}{0,151,1}
\definecolor{myr}{RGB}{234,0,0}

\newtheorem{problem}{Problem}

\begin{document}
%

    \title{Shape Uncertainty Quantification for Electromagnetic Wave Scattering via First-Order Sparse Boundary Element Approximation}
%
%
%

\author{Paul Escapil-Inchausp\'e, and Carlos~Jerez-Hanckes,~\IEEEmembership{Member,~IEEE}
\thanks{This work was supported in part by ``Beca Postdoctoral Fondecyt 3230088" and ``Fondecyt Regular 1231112" (\emph{Corresponding author: Carlos~Jerez-Hanckes}).} 
\thanks{Carlos~Jerez-Hanckes and P.~Escapil-Inchausp\'e are  with the Faculty of Engineering and Sciences, Universidad Adolfo Ibáñez, Santiago, Chile. The latter is also with the Data Observatory Foundation, Santiago, Chile (e-mail: carlos.jerez@uai.cl).}
}

%
%

\markboth{}
{Escapil-Inchausp\'e \& Jerez-Hanckes: Shape UQ for EM wave scattering via FOSB approximation}

%



\maketitle

\begin{abstract}
Quantifying the effects on electromagnetic waves scattered by objects of uncertain shape is key for robust design, particularly in high precision applications. Assuming small random perturbations departing from a nominal domain, the first-order sparse boundary element method (FOSB) has been proven to directly compute statistical moments with poly-logarithmic complexity for a prescribed accuracy, without resorting to computationally intense Monte Carlo simulations. However, implementing the FOSB is not straightforward. To this end, we introduce an easy-to-use with open-source framework to directly apply the technique when dealing with complex objects. Exhaustive computational experiments confirm our claims and  demonstrate the technique's applicability as well as provide pathways for further improvement.
\end{abstract}

\begin{IEEEkeywords}
Electromagnetic Wave Scattering, Uncertainty Quantification, Shape Derivative, Combination Technique, Boundary Element Methods
\end{IEEEkeywords}

%
\IEEEpeerreviewmaketitle

\section{Introduction}
\label{sec:intro}

\IEEEPARstart{U}NCERTAINTY Quantification (UQ) is a well-established research field comprising the characterization and estimation of randomness in applied problems \cite{sullivan2015introduction}. In the case of computational electromagnetism (CEM), one assumes aleatoric UQ, i.e.~Maxwell equations hold at all times and randomness is due to parameters, shapes and sources. In particular, we aim at quantifying the effects of random shape perturbations from a nominal domain on EM fields, i.e.~perform shape UQ. Such models pertain, for instance, to radar or radio-telescope mirror robust design as the slightest stochastic deformation---due either manufacturing or operation conditions---from the original shape is likely to produce undesirable effects. Thus, given a class of perturbed domains defined over a probability space $\Omega$, shape UQ seeks to estimate the mean and higher-order statistical moments for random EM fields in those domains. This is key, for example, to carry out sensitivity analyses as the covariance and its diagonal terms (variance) provide confidence intervals.

Computationally, one can resort to different strategies to estimate the above. For instance, many efforts have been made to devise reliable, non-intrusive methods to account for randomness. Amongst them, one finds Monte Carlo (MC) simulations and variations such as Multi-Level or Quasi MC \cite{caflisch_1998MCQMC}. Approximation of parametric uncertainties can also be accelerated via sparse grids \cite{bungartz_griebel_2004} or Smolyak quadrature \cite{zechsmolyak}. However, these methods entail carrying out full-fledged simulations for multiple realizations, rendering the approach prohibitive for 3D CEM due to the large number of degrees of freedom (dofs) inherently required.

Intrusive methods are generally computationally cheaper than the above and depend on how the random perturbations behave, leveraging prior knowledge of a nominal shape. For instance, when considering large domain deformations, existing shape holomorphy results imply provable convergence rates for domain-to-solution mapping approximations \cite{MaxwellShapeHolomorphy}, with rapid computation via finite element for bounded domains \cite{DUQaylwin,MLDUQaylwin}, or by deep learning surrogates evaluation \cite{SchwabZechDeepLearning, ScarabosioDNN}. However, the efficiency of the method depends on the smoothness of the underlying random perturbation, e.g., the coefficients' decay rate for parametric random transformations. 

Alternatively, for small magnitude shape (random) deformations one could turn to perturbation methods such as first-order approximations (FOAs) that hinge on the concept of shape derivative (SD). These methods allow for linearizing the problem for each domain realization and transferring the stochasticity to the boundary data. Under suitable smoothness requirements, the SD can be recast as a well-posed boundary value problem (BVP) \cite{hettlichdomainderivativeEM}. For more details and applications of the SD, we refer to \cite{costabel2012shape,Hiptmair_2018SDii,colton1998inverse,Hagemann_2019}.

In the present work, we consider EM wave scattering problems by perfect electric conductor (PEC) and dielectric (DE) obstacles in open space. Since the domain is unbounded, we solve them employing boundary integral equations (BIEs) \cite{Nedelec} discretized by the boundary element methods (BEM) or Rao-Wilton-Glisson elements \cite{BuffaHiptmair2003}.  As we assume small perturbations, we can perform FOA wherein the resulting SD is also solution of a BIE over which we take statistical moments. This leads to a deterministic tensor operator equation, which is then discretized via BEM. As such, it is often useful to use sparse tensor approximations \cite{vonPetersdorff2006} or the combination technique (CT) \cite{HARBRECHT2013128kthCt} to reduce the dimensionality of the resulting tensor equation. This so-called first-order sparse boundary element method (FOSB) was introduced for Laplace in both bounded and unbounded domains in \cite{sparse3} and \cite{CHERNOV20151401}, respectively. Later, it was extended to Maxwell \cite{JS15_613} and Helmholtz \cite{FOSB_Helmholtz,8169061} scattering. The tensor equation can also be reduced to a low-rank approximation \cite{Harbrecht2014Robin,dolzhighorder} through pivoted Cholesky decomposition \cite{HARBRECHT2012lr}. Also, the right-hand side can be compressed via $\mH$-matrix representations \cite{bebendorf2008hierarchical,dolzhighorder}. Recently, tensor operator equations where solved using the novel physics-informed neural networks \cite{escapil2022physics}. 

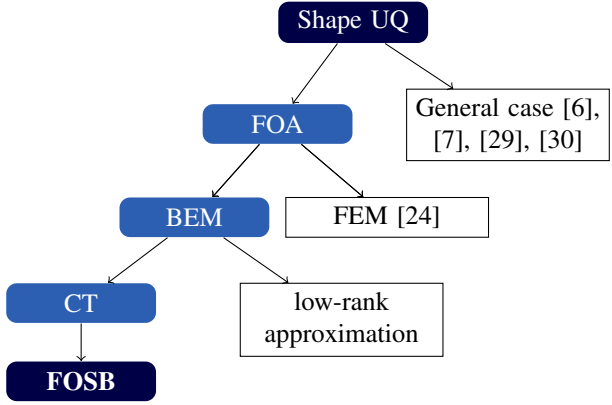
\begin{figure}[t]
 \centering
  \begin{tikzpicture}[paths/.style={->, thick, >=stealth'}, node distance = 2cm, auto]
   \node [cloud] (init) {\textcolor{white}{Shape UQ}};
   \node [back_b, below left = 0.8cm and -1cm of init] (01) {\textcolor{white}{FOA}};
   \node [back, below right = 0.6cm and -0.3cm of init] (00) {General case \cite{hiptmair2018large,harbrecht2019rapid,DUQaylwin,MLDUQaylwin}};

   \node [back_b, below left = 0.7cm and -0.9cm of 01] (10) {\textcolor{white}{BEM}};
   \node [back, below right = 0.7cm and -.9cm of 01] (11) {FEM \cite{Harbrecht2014Robin}};

   \node [back_b, below left = 0.6cm and -0.5cm of 10] (20) {\textcolor{white}{CT}};
   \node [back, below right = 0.6cm and -0.4cm of 10] (21) {low-rank approximation};

  \node [back_g, font=\bf, below = 0.5cm and -0.5cm of 20] (30) {\textcolor{white}{FOSB}};

   \draw[->] (init) -- (00) ;
   \draw[->] (init) -- (01) ;

   \draw[->] (01) -- (10) ;
   \draw[->] (01) -- (11) ;

   \draw[->] (01) -- (10) ;
   \draw[->] (01) -- (11) ;

   \draw[->] (10) -- (20) ;
   \draw[->] (10) -- (21) ; 
   \draw[->] (20) -- (30) ;
\end{tikzpicture}
\caption{Background of the FOSB method (in blue) and nomenclature for alternative approaches.}
\label{picture:background}
\end{figure}

In a nutshell, the FOSB consists in the following steps:
\begin{enumerate}
\item[I.] Apply the FOA to linearize the random problem;
\item[II.] Reduce the problem via boundary integral representation and take statistical moments to obtain a boundary tensor operator equation;
\item[III.] Numerically solve the underlying tensor equation via BEM coupled with the combination technique (CT).
\end{enumerate}
We sum up the background for shape UQ and FOSB method in \Cref{picture:background}. 
The mathematical tools for the FOSB for Maxwell scattering were provided in \cite{JS15_613}. For an introduction to the FOSB methodology, we refer the reader to the recent survey \cite{FOSB_Helmholtz} for the Helmholtz scattering problem. Furthermore, the use of dual basis makes it possible to guarantee stable $\bL^2(\Gamma)$-pairings and accurately evaluate the right-hand side for the SD \cite{Hagemann_2019}. Still, the numerical implementation of FOSB can be challenging and though exhaustive numerical experiments were presented in \cite{FOSB_Helmholtz} for the Helmholtz case, to our knowledge, no actual FOSB EM code has been implemented.

In this note, we present the second moment FOSB for PEC and DE scatterers in an intuitive format geared towards EM practitioners. To assess its feasibility and efficacy, we subject it to extensive numerical investigations using the novel Bempp-UQ package\footnote{https://github.com/pescap/bempp-uq/} ensuring also compliance with FAIR principales (Findability, Accessibility, Interoperability, and Reusability) for scientific data management and stewardship \cite{wilkinson2016fair}. Our practical findings are (i) CT allows to reduce computational requirements of its full tensor variant by a factor ranging between $5$ and $100$; (ii) FOA supplies an accurate approximation to the variance considered; (iii) FOSB is robust and outperforms MC by a factor of $5.87$ (resp.~$7.44$) in terms of dofs (resp.~total execution time) for a complex case in \Cref{subsec:Complex_case}.

After setting the problem in \Cref{sec:Problem setting}, we provide FOA in \Cref{sec:FOA} and discretizations \Cref{sec:Galerkin}. Complete numerical experiments in \Cref{sec:Numexp} confirm our claims and further research avenues are presented in \Cref{sec:conclu}.

\section{Problem setting}
\label{sec:Problem setting}
\subsection{Notation and Definitions}
\label{subsec:Notations}
Throughout this work, we refer EM time-harmonic wave problems as $(P_\beta)$ with $\beta=0$ for PEC and $\beta=1$ for DE. We assume an angular frequency $\omega:= 2 \pi f$ for a time dependence $\e^{-\imath \omega t}$, $f > 0 $ and $\imath^2 = -1$. Let $D \subset \IR^3$ be an open bounded Lipschitz domain with boundary $\Gamma :=\partial D$, exterior unit normal $\bn$, and complement $D^c:=\IR^3 \backslash \overline{D}$. Define $\fD := D$ for $\beta = 0$ and $\fD:= D^c \cup D$ for $\beta=1$ along with  $D^c=D^0$ and $D^c=D^1$. We set each medium with permittivity and permeability $\upmu_i,\epsilon_i$ along with wavenumbers $k_i := \omega \sqrt{\upmu_i \epsilon_i}$, $i=0,1$, with $k : = k_0$ and wavelength $\lambda := \frac{2 \pi}{k}$. Furthermore, let 
\be \label{eq:relativeParameters}
\epsilon_r :=\frac{\epsilon_1}{\epsilon_0} , \quad \upmu_r :=\frac{\upmu_1}{\upmu_0} \quad \text{and} \quad \eta := \sqrt{\frac{\upmu_r}{\epsilon_r}}.
\ee
For any $s \geq 0 $, we introduce the standard Sobolev spaces $\bH^s(D)$ (resp.~$\bH(\Gamma)$) with $\bL^2(D):= \bH^0(D)$ (resp.~$\bL^2(\Gamma) := \bH^0(\Gamma)$) \cite{BuffaHiptmair2003}. Within the unbounded $D^c$, we use the $\text{loc}$-subscript for spaces with bounded Sobolev norm over each compact subset $K \Subset D^c$. For vectors, $\cdot$ and $|\cdot|$ denote the Euclidean product and norm, respectively. We recall the separable Hilbert space:
\be  \bH(\bcurl , D) :=  \{\bU \in \bL^2 (D) | \bcurl \bU \in \bL^2(D)\}
\ee
with $\bcurl$ being the curl operator. For any smooth vector field $\bU$, we introduce electric and magnetic traces \cite{BuffaHiptmair2003}:
\be
\bgamma_D \bU : = \bU|_\Gamma \times \bn \quad \text{and} \quad \bgamma_N\bU := \frac{1}{\imath k} \bgamma_D(\bcurl \bU|_\Gamma).
\ee
Accordingly, we introduce the twisted tangential trace:
\be 
 \bgamma_T \bU := \bn \times \bgamma_D \bU.
\ee 
Finally, we define the space
\be\label{eq:XGammaspace}
 X(\Gamma) \equiv \bH_\times^{-1/2}(\divg_\Gamma, \Gamma)
\ee 
of tangential traces on the boundary such that the mapping $\bgamma_D : \bH(\bcurl, D) \to X(\Gamma) $ is continuous and surjective, with $\divg_\Gamma$ being the tangential divergence operator \cite{BuffaHiptmair2003}. For smooth boundaries and any $s\geq 0$, let smoothness spaces \cite[Section 7.2]{JS15_613}:
 \be\label{eq:SmoothnesXs} 
 X^s(\Gamma) :=  \bH^{-1/2 +s } (\divg_\Gamma , \Gamma).
\ee 
For $s \geq 0 $, we signal the natural trace space for each problem $(P_\beta)$ as 
\be\label{eq:XBetaGammaspace}
X^s_\beta(\Gamma) := \begin{cases} X^s(\Gamma) \quad \text{for} \quad \beta = 0,\\
X^s(\Gamma)^2 \quad \text{for} \quad \beta = 1.
\end{cases}
\ee
For $\bu, \bv \in X(\Gamma)$ the duality pairing for tangential traces reads 
\be 
\langle\bu, \bv  \rangle_\times  := \int_\Gamma \bu \cdot (\bn \times \bv )\mathrm{d}\Gamma.
\ee
Naturally, we extend the duality product to smooth tangential pairs as:
\be 
 \left\langle \begin{pmatrix} \bu \\ \bp \end{pmatrix} , \begin{pmatrix}\bv \\ \bq\end{pmatrix} \right\rangle_\times  : = \langle \bu,\bv\rangle_\times + \langle \bp , \bq\rangle_\times .
\ee
Traces with respect to $D^c$ are denoted by $c$-superscript. Then, we define trace jumps and averages for $\cdot  \in \{D,N\}$ as
\be 
[\bgamma_\cdot]_\Gamma := \bgamma^c_\cdot - \bgamma_\cdot \quad \text{and} \quad \{ \bgamma_\cdot  \}_\Gamma := \frac{1}{2}(\bgamma^c_\cdot + \bgamma_\cdot) .
\ee
Let $(\Omega, \mA, \IP)$ be a probability space on a separable Hilbert space $X$. For a random field $\bU : \Omega \to X$, we introduce the first and second statistical moments as well as the variance:
\begin{align}\label{eq:stat}
\IE[\bU(\omega)]& : = \int_\Omega \bU (\bx,\omega)d \IP(\omega),\\
\mM^2[\bU(\omega)] &: = \int_\Omega \bU (\bx_1,\omega) \overline{\bU}(\bx_2,\omega) d \IP (\omega),\\
\IV [\bU(\omega)] & := \mM^2\left[\bU(\omega) -\IE[\bU(\omega)] \right]|_{\bx_1 = \bx_2}.
\end{align}
Finally, we set $X^{(2)} := X \otimes X$ and introduce the tensor duality product $\langle \cdot, \cdot \rangle_\times^{(2)}$ \cite[Section 6.2]{JS15_613}.
\subsection{Boundary Integral Operators and Equations}
\label{subsec:IntegralOperators}
We now introduce boundary potentials and boundary integral operators (BIOs). The electric and magnetic potentials are defined as:
\be \label{eq:potential_operators}
\begin{split}
\mE_k \bv (\bx )& : = \imath k \int_\Gamma \bv ( \by) G_k (\bx, \by) \mathrm{d}\Gamma (\by) \\
 & - \frac{1}{\imath k} \nabla_\bx \int_\Gamma  \divg_\by ( \bv(\by) ) G_k(\bx,\by) \mathrm{d}\Gamma (\by), \\
 \mH_k \bv (\bx ) & : = \bcurl \int_\Gamma \bv ( \by) G_k (\bx, \by) \mathrm{d}\Gamma (\by) ,
\end{split}
\ee 
wherein $G_k(\bx, \by) = \dfrac{\exp ( \imath k |\bx - \by| )}{4 \pi |\bx - \by| }$ and $\divg$ is the divergence operator. Next, we introduce the BIOs $X(\Gamma) \to X(\Gamma)$ with $X(\Gamma)$ as in \eqref{eq:XGammaspace}:
\be \label{eq:boundary_operators}
\begin{cases}
\text{Identity:} & \mI , \\
\text{EFIO:} & \mT_k  : = \{ \bgamma_D \mE_k \}_\Gamma  = - \{ \bgamma_N \mH_k\}_\Gamma , \\
\text{MFIO:} & \mK_k : = \{ \bgamma_D \mH_k \} = \{ \bgamma_N \mE_k\}_\Gamma. \\
\end{cases}
\ee 
We also introduce the multitrace identity and its scaled version
\be \label{eq:BOmultitraceI}
\bmI : = \begin{pmatrix} \mI &0 \\ 0 & \mI  \end{pmatrix}, \quad \hbmS : = \begin{pmatrix}  \epsilon_r^{-1/2 }\mI   & 0 \\ 0  & \upmu_r^{-1/2}\mI  \end{pmatrix}
\ee 
along with the multitrace operator and its scaled counterpart:
\be \label{eq:BOmultitraceA}
 \bmA_k : = \begin{pmatrix} \mK_k  & \mT_k  \\ -\mT_k  & \mK_k \end{pmatrix} , \quad  \hbmA_k : = \begin{pmatrix}  \mK_k  & \eta \mT_k   \\ - \eta^{-1} \mT_k  & \mK_k \end{pmatrix}.
\ee 
\subsection{Deterministic Scattering Problems}\label{subssec:Deterministic}Consider an incident field $\bE^{\text{inc}}$ such that $\bcurl \bcurl \bE^{\text{inc}} - k^2 \bE^{\text{inc}} = 0 $ in $D^c$, and denote by SM$(\cdot)$ the Silver-M\"uller radiation condition \cite[Eq.~2.3]{JS15_613}. We are now ready to introduce the mathematical formulation for $(P_\beta)$.
\begin{problem}[$(P_0)$: PEC]
Seek the total field $\bE =\bE^{\text{inc}} + \bE^{\text{sc}} \in \bH_\text{loc}(\bcurl,D^c)$, such that
\be
\begin{array}{rll}
\bcurl \bcurl \bE - k^2 \bE = & 0  &  \text{ in } D^c,\\
\bgamma_D^c \bE   = &  0 & \text{ on }\Gamma,\\
\textup{SM}(\bE^\textup{sc})& &\text{ for }|\bx| \to \infty.
\end{array}
\ee 
\end{problem}
The induced electrical current density $\bj := \gamma_N^c \bE \in X(\Gamma) $ satisfies the EFIE (Electric Field Integral Equation):
\be\label{eq:pec}\mT_k (\bj) =  \bgamma_D^c \bE^\text{inc},
\ee
with $\mT_k$ in \eqref{eq:boundary_operators}. We assume that $k^2$ is not an interior electric eigenvalue, ensuring well-posedness of the EFIE \cite[Remark 2.4]{JS15_613}. Furthermore, fields can be recovered via the Stratton-Chu representation formula in $D^c$ \cite[Eq.~(9)]{Hagemann_2019}:
\be
\bE = \bE^\textup{inc}- \mE_k (\bj),
\ee
with $\mE_k$ as in \eqref{eq:potential_operators}.

\begin{problem}[$(P_1)$: DE]
Seek the total field $(\bE^0,\bE^1) =(\bE^{\text{inc}} + \bE^{\text{sc}}, \bE^1) \in \bH_\text{loc}(\bcurl, D^0) \times \bH(\bcurl,D^1)$, such that
$$
\begin{array}{rll}
\bcurl \bcurl \bE^i - k_i^2 \bE^i &=  0  &\text{in} \quad  D^i,~i=0,1,\\
 \epsilon_0^{-1/2} \bgamma^c_D  \bE^0 &= \epsilon_1^{-1/2} \bgamma_D \bE^1   & \text{on} \quad \Gamma,\\
 \upmu^{-1/2}_0 \bgamma^c_N \bE^0   &= \upmu^{-1/2}_1 \bgamma_N \bE^1  &  \text{on} \quad \Gamma, \\[4pt]
 \textup{SM}(\bE^\textup{sc}) & \text{for}\quad |\bx| \to \infty.
 \end{array}
 $$
\end{problem}
Moreover, we introduce
$$
\bxi^\text{sc}  \equiv \begin{bmatrix} \bj^\text{sc} \\ \bm^\text{sc} \end{bmatrix} := \begin{bmatrix} \bgamma_D^c \bE^\text{sc}\\ \bgamma_N^c \bE^\text{sc} \end{bmatrix},
$$
and similarly for $\bxi^\text{inc}$ and $\bxi^1$. Consequently, it holds that $\bxi^\text{sc} \in X(\Gamma)^2$ yields the PMCHWT equation \cite[Eq.~(12)]{Hagemann_2019}:
\be \label{eq:pmchwt}
(\bmA_0 + \hbmA_1 )\bxi^\text{sc} = \left( \frac{1}{2} \bmI - \bmA_1\right) \bxi^\text{inc} =:\bff
\ee
with $\bmA_i\equiv \bmA_{k_i}$, $i=0,1$, BIOs defined in \eqref{eq:BOmultitraceI} and \eqref{eq:BOmultitraceA}. As for the PEC case, the Stratton-Chu representation formulae read
\begin{align*}
\bE^0  &= \bE^\textup{inc}- \mE_0 \bj^\text{sc} - \mH_0 \bm^\text{sc}\textup{, and}\\
\bE^1 & = - \mE_1 \bj^1 - \mH_1 \bm^1.
\end{align*}
with potential sub-indices $0,1$ referring to wavenumbers $k_0$ and $k_1$, accordingly. By introducing $\bmR_i:=(-\mH_i,-\mE_i)$, for $i=0,1$, there holds that
\be
\bE = \bE^\textup{inc} + \bmR(\bxi).
\ee
For both PEC and DE problems, the electric far-field $\bF$ is for $|\bx| \to \infty$ such that 
\be
\bE^\text{sc} (\bx)   = \frac{e^{\imath k |\bx|}}{4 \pi |\bx|} \left[ \bF \left( \frac{\bx}{|\bx|}\right) + \mO \left( \frac{1}{|\bx|}\right)\right].
\ee
At $z=0$, the far-field reads in Cartesian coordinates as 
\be 
\label{eq:FFcomp}\bF (\theta) =  F_x (\theta) \hat{\bee}_x + F_y (\theta) \hat{\bee}_y + F_z (\theta)\hat{\bee}_z
\ee
for $\theta:= \text{atan2}(y,x) \in [0, 2\pi]$. We set the radar cross section (RCS) in decibels as: 
\be \label{eq:RCS}
\text{RCS}(\bF)(\theta) := 10 \log_{10} \left(4 \pi \frac{|\bF(\theta)|^2}{|\bF^\text{inc}(\theta)|^2} \right) .
\ee
In our analysis, we shall study the RCS for each FF component in \eqref{eq:FFcomp}:
\be 
\text{RCS}_\cdot (\theta) :=  \text{RCS} (F_\cdot)(\theta) \quad \text{for} \quad \cdot \in \{x,y,z\}.
\ee 
\section{FOA method}
\label{sec:FOA}
\begin{figure}[t]
\center
\includegraphics[width=.77\linewidth]{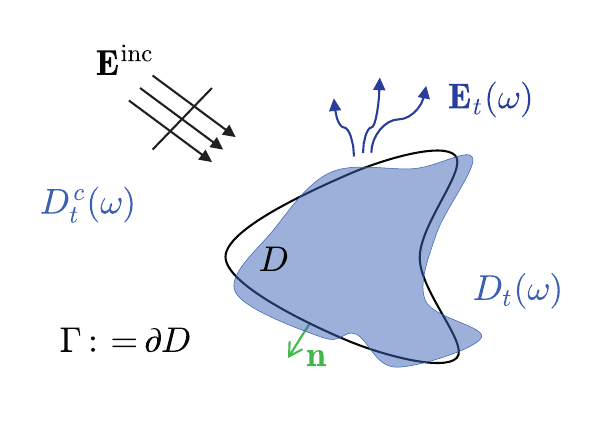}
\caption{PEC problem setting with $D^c$ the nominal exterior domain with boundary $\Gamma$ (dark) and exterior normal $\bn$ (green). $D_t^c(\omega)$ is a realization of the random exterior domain (blue).}
\label{fig:problem}
\end{figure}

Now, let $(\Omega,\mA,\IP)$ be a complete probability space. Following \cite[Section 2.3]{FOSB_Helmholtz}, we consider a centered random velocity field $\bv\in L^2(\Omega,\IP;\boldsymbol{C}^2(\Gamma;\IR^3))$ such that $\IE[\bv (\omega)] = 0 $ and $\|\bv(\omega) \|_{\boldsymbol{C}^2(\Gamma;\IR^3) } \leq C $ uniformly for all $\omega \in \Omega$. We consider a nominal bounded domain $D$ of class $C^2$, and introduce a family of random surfaces $\{\Gamma_t\}_t$ for any $ |t| \ll 1$ via the mapping:
\be
\Omega \ni \omega \mapsto \Gamma_t (\omega) = \{\bx + t \bv(\omega), \bx \in \Gamma \} .
\ee
Therefore, for small $|t|\ll 1$, the interior of $\Gamma_t$ family defines $\IP$-a.s.~Lipschitz domains $\{D_t\}_t$. Furthermore, we introduce $\bE_t(\omega)$ the solution of $(P_\beta)$ for each realization over the domain $\fD_t (\omega)$ with $\fD_t(\omega)=D^c_t(\omega)$ for $\beta=0$ and similarly when $\beta=1$. In \Cref{fig:problem} above, we represent the PEC problem.

We aim at quantifying the statistical moments of $\bE_t(\omega)$ (see \eqref{eq:stat}) for small enough $|t| \ll 1$. This is conducted through shape calculus. Following \cite[Section 2.1]{DOLZ2018506}, we introduce the compact set $\fK$ such that:
\be \label{eq:Kcompact}
\fK \Subset \mathsf{D}_t^{\cap \Omega}, \quad \mathsf{D}_t^{\cap \Omega} : = \bigcap_{\omega \in \Omega} \mathsf{D}_t(\omega). 
\ee 

Let $\bE'(\omega)$ denote the SD of $\bE$ for the scattering problem $(P_\beta)$ on the nominal domain but with stochastic boundary data. With it, one can write two statistical moments for the electric fields in $(P_\beta)$  as follows \cite{JS15_613}:
\begin{align*}
\IE [\bE_t(\omega) ] &= \bE + \mO(t^2) ~\text{in}~\bH(\bcurl, \fK),\\
\mM^2 [\bE_t(\omega) - \bE] & = t^2 \mM^2[\bE'(\omega)] + \mO(t^{3}) ~\text{in} ~ \bH(\bcurl,\fK)^{(2)}.
\end{align*}
Therefore, FOA amounts to computing quantities defined on the nominal domain as being: (i) $\bE$ the solution to the unperturbed scattering problem; and (ii) $\mM^2[\bE'(\omega)]$, which is recast as a tensor operator equation with stochastic right-hand side. Moreover, it holds that
\be 
\IV [\bE_t (\omega)]  = t^2 \IV[\bE'(\omega)] + \mO(t^3) \quad \text{in} \quad \bH (\bcurl, \fK).
\ee
Consider first the PEC case and set $\text{d} \bj'(\omega)$ as the electric trace of $\bE'(\omega)$, i.e.~$\text{d}\bj'(\omega):= \gamma_N^c \bE'(\omega) \in X(\Gamma) $. For each realization $\omega \in \Omega$, there holds that \cite[Eq.~20]{hettlichdomainderivativeEM}: 
\be
\begin{split}
\mT_k ( \text{d} \bj' (\omega)) &= \nabla_\Gamma [ v_n (\omega) E_n] \times \bn - \imath k v_n (\omega) \gamma_T^c \mathbf{H} \\
 &  = : \bg_1 (\omega)+ \bg_2(\omega) \equiv \bg(\omega)
\end{split}
\ee
with $\mathbf{H}$ the associated magnetic field, $E_n := \bE \cdot \bn$ and $v_n(\omega) := \bv(\omega) \cdot \bn$. Consequently, application of the second statistical moment yields
\be \label{eq:secondPEC}
\begin{split}
(\mT_k
 \otimes \overline{\mT_k}) \mM^2[\text{d} \bj' (\omega)]& = \mM^2[\bg(\omega)]\\
& = \IE[\bg_1 \otimes \overline{\bg}_1] +  \IE[\bg_1 \otimes \overline{\bg}_2] \\
& + \IE[\bg_2 \otimes \overline{\bg}_1] + \IE[\bg_2 \otimes \overline{\bg}_2]
\end{split} 
\ee 
wherein
\be\nonumber
\begin{split}
\IE[\bg_1 \otimes \overline{\bg}_2] & = (\nabla_\Gamma \otimes \nabla_\Gamma) (\mM^2[v_n(\omega)] (E_n \otimes \overline{E_n})) (\times \bn  \otimes \times \bn),\\
 \IE[\bg_2 \otimes \overline{\bg}_1] & = -\imath k (\nabla_\Gamma \otimes \mI) (\mM^2 [v_n(\omega)]  (E_n \otimes \bgamma_T^c \overline{\mathbf{H}})),\\
 \IE[\bg_1 \otimes \overline{\bg}_2] & = -\imath k (\mI \otimes \nabla_\Gamma) (\mM^2 [v_n(\omega)] ( \bgamma_T^c \mathbf{H} \otimes \overline{E_n} )),\\
 \IE[\bg_2 \otimes \overline{\bg}_2] & = - k^2 \mM^2[v_n(\omega)] (\bgamma_T \mathbf{H} \otimes \bgamma_T^c \overline{\mathbf{H}} ) .
\end{split}
\ee 
As a consequence, $\mM^2[\bE'(\omega)])$ can be built as
$$
\mM^2[\bE'(\omega)] : = \mE_k^{(2)} \mM^2[\text{d} \bj'(\omega)]
$$
with $\mM^2[\text{d} \bj'(\omega)]$ in \eqref{eq:secondPEC}. 

For the DE case \cite[Eq.~(24)]{Hagemann_2019}, it holds that for each $\omega \in \Omega$:
\be
\begin{split}
\label{eq:PMCHWT_SD_omega}
\bh(\omega) :=( \bmA_0 + \hbmA_1 )\text{d} \bxi'(\omega) &=\!  \left(  \frac{1}{2} \bmI - \hbmA_1 \right)\!\! \begin{pmatrix} \bh_1^c (\omega) \\ \bh_2^c (\omega)\end{pmatrix} \\
& +\! \hbmS\! \left( \hbmA_0 - \frac{1}{2} \bmI  \right)   \!\!\begin{pmatrix} \bh_1(\omega)  \\ \bh_2 (\omega)\end{pmatrix}
 \end{split}
\ee
with
$$
\begin{pmatrix} \bh_1^c (\omega) \\ \bh_2^c(\omega) \end{pmatrix} : = \begin{pmatrix} \nabla_\Gamma [ v_n (\omega) E_n] \times \bn - \imath k_0 v_n (\omega) \gamma_T^c  \mathbf{H}^0  \\ \nabla_\Gamma [ v_n (\omega) E_n] \times \bn + \imath k_0 v_n(\omega)  \gamma_T^c \mathbf{H}^0\end{pmatrix}
$$
and
$$
\begin{pmatrix} \bh_1 (\omega) \\ \bh_2 (\omega)\end{pmatrix} : = \begin{pmatrix} \nabla_\Gamma [ v_n (\omega) E_n] \times \bn - \imath k_1 v_n  (\omega)\gamma_T \mathbf{H}^1  \\ \nabla_\Gamma [ v_n (\omega) E_n] \times \bn + \imath k_1 v_n (\omega) \gamma_T \mathbf{H}^1 \end{pmatrix}.
$$
Therefore:
\be \label{eq:de_first}
( \bmA_0 + \hbmA_1 )\otimes(\overline{ \bmA_0 +\hbmA_1} ) \mM^2[\text{d} \bxi '(\omega) ]=   \mM^2[\bh(\omega)]
\ee 
where $\text{d} \bxi '(\omega)\in X(\Gamma)^2$ is the trace of the SD of the DE problem solution pair, and
\be 
\mM^2 \left[\begin{pmatrix}\bE^{0'}\\ \bE^{1'}\end{pmatrix}(\omega)\right] : = \mR^{(2)} \mM^2 [d\bxi'(\omega)]. 
\ee 
\section{Galerkin Method}
\label{sec:Galerkin}
We set a maximum mesh refinement level $L>0$. For $l=0,\ldots,L$, we generate a family of meshes $\Gamma_{h_l}$ with mesh-width $h_l>0$, representing $\Gamma$, consisting of a subdivision  of $\Gamma$ into a set of planar triangular non-overlapping elements. The number of points per wavelength---precision---is
\be\label{eq:precision} 
r_l : = \frac{2 \pi}{h_l  k_0} = \frac{\lambda}{h_l}. 
\ee
We assume that $h_l \sim q^{-l}$ with $q>1$, e.g. $q\in \{2,4\}$ \cite[Eq.~3.1]{hiptmair2013sparse}. We discretize the BIEs by approximating the unknowns using $\divg$-conforming functions with discrete domain and test spaces, $X_l^\text{dom}$ and $X_l^\text{test}$, respectively, being finite dimensional subpaces of $X(\Gamma)$ of dimension $N_l$. For details concerning the stable dual pairing, refer to \cite[Section 3.2]{KLEANTHOUS2022111099}. Also, we forgo the error analysis incurred by non-conforming meshes \cite{Curved}.

\subsection{First and Second Full Moments}
\label{subsec:First}
For the first moment\revn{s}, the Galerkin discretization of the EFIE \eqref{eq:pec} in case of PEC leads to approximations of $\bj$ by $\bj_L \in X_L^\text{dom}$ solutions of
$$
\left\langle \mT_k \bj_L   , \bphi_L \right\rangle_\times = \left\langle \bgamma_D^c \bE^\text{inc},  \bphi_L \right\rangle_\times\quad \forall \bphi_L \in X_L^\textup{test}.
$$
For the PMCHWT \eqref{eq:pmchwt} approximations we solve for $\bxi^\text{sc}_L \in (X_L^\text{dom})^2$ the system
$$
\left\langle (\bmA_0 + \hbmA_1 )\bxi^\text{sc}_L , \bphi_L \right\rangle_\times = \left\langle \bff, \bphi_L \right\rangle_\times\quad \forall \bphi_L \in (X_L^\textup{test})^2.
$$ 
Both linear systems are well posed and of the form:
\be
\label{eq:linear_system}
\bZ_L\bu_L = \bb_L .
\ee
For each $\bu_L$ one can associate an element $u_L\in (X_L^\text{dom})^{\beta+1} $. Assume that $ u \in X^s_\beta (\Gamma)$  for any $s > 0$. Then, there exists $L_0(k;\beta) \in\mathbb{N} $ such that for all $L \geq L_0$, the Galerkin approximation $u_L$ satisfies \cite[Theorems 5.7 and 5.10]{JS15_613}
\be 
\| u - u_L \|_{X_\beta(\Gamma)} \leq C h^s_L \| u\|_{X_\beta^s(\Gamma)}.
\ee
Similar considerations hold for second moment. For the PEC case \eqref{eq:secondPEC}, we seek $\Sigma_L \in (X_L^\text{dom})^{(2)}$ such that
$$
\left\langle (\mT_k \otimes \overline{\mT_k})  \Sigma_L , \bphi_L \right\rangle_\times^{(2)} = \left\langle \mM^2[\bg] ,  \bphi_L \right\rangle_\times^{(2)} \ \forall \bphi_L \in (X_L^\textup{test})^{(2)}.
$$
For the DE case \eqref{eq:de_first}, we look for $\Sigma_L \in \left((X_L^\text{dom})^{2}\right)^{(2)}$ such that: 
\begin{align*}
\left\langle ( \bmA_0 + \hbmA_1 )\otimes(\overline{ \bmA_0 +\hbmA_1} )  \Sigma_L , \bphi_L \right\rangle_\times^{(2)} &= \left\langle  \mM^2[\bh], \bphi_L \right\rangle_\times^{(2)} \\
& \forall \bphi_L \in ((X_L^\textup{test})^2)^{(2)} .
\end{align*}
Again, these variational formulations lead to well-posed linear systems of the form:
\be 
\label{eq:linear_system2}
 (\bZ_L\otimes \overline{\bZ}_L) \boldsymbol{\Sigma}_L = {\bf C}_L.
\ee
As shown in \cite[Theorem 6.3]{JS15_613}, assuming that $\Sigma$ the unknown belongs to $X^s_\beta (\Gamma)^{(2)}$ for any $s\geq0$, there exists $L_0(k;\beta)\in\mathbb{N}$ such that for all $L \geq L_0$, the Galerkin approximation $\Sigma_L \in ((X_L^\text{dom})^2)^{(2)}$ to $\Sigma$ yields  :
\be \label{eq:errorSecond}
\|  \Sigma - \Sigma_L \|_{X_\beta(\Gamma)^{(2)}} \leq C h^s_L \| \Sigma\|_{X_\beta^s(\Gamma)^{(2)} }.
\ee
\subsection{Combination Technique}
\label{subsec:CT}
We apply the CT to approximate second order tensor operator BIEs. It allows for paralellization and is non-intrusive for BEM solvers. Let us introduce the indices:
\begin{align*}
\Lambda_+(L_0, L) &: = \{ (l_1, l_2)  : \ l_1 + l_2 = L + L_0 \} , \\
\Lambda_- (L_0, L)& : = \{ (l_1, l_2) : \ l_1 + l_2 = L + L_0 - 1\},\\
\Lambda( L_0,L)  &: = \Lambda_+(L_0, L)  \cup \Lambda_-(L_0, L).
\end{align*}    
The combination technique amounts to solving:
\be \label{eq:CTsubsystem}
(\bZ_{l_1} \otimes \overline{\bZ}_{l_2})  \boldsymbol{\Sigma}_{l_1,l_2} = {\bf C}_{l_1,l_2} 
\ee
for $\lambda \in \Lambda$, yielding the sparse tensor approximation:
\be
\boldsymbol{\hat{\Sigma}}_L : =\sum_{(l_1, l_2) \in \Lambda_+} \boldsymbol{\Sigma}_{l_1, l_2} - \sum_{(l_1, l_2) \in \Lambda_-} \boldsymbol{\Sigma}_{l_1, l_2} .
\ee
As observed in \cite[Section 6.1]{FOSB_Helmholtz}, for Hermitian right-hand sides, $\bSigma_{l_1,l_2} = \bSigma_{l_2,l_1}^H$ for any $(l_1,l_2) \in \Lambda$. Consequently, the indices in $\Lambda$ reduce to:
$$
\Lambda_H(L_0,L) : = \Lambda(L_0,L) \cup \{ (l_1, l_2 ) \ :\ l_1 \leq l_2 \}
$$
thereby avoiding computing half of the sub-blocks \cite{FOSB_Helmholtz}. The number of dofs for each sub-block is $N_{l_1,l_2} : = N_{l_1} \cdot N_{l_2}$. Then,  total number of dofs for the CT is
\be 
\hat{N}_L: = \sum_{(l_1,l_2) \in \Lambda} N_{l_1,l_2}
\ee
and the maximum sub-block size reads:
\be 
\hat{N}_L^\text{max} : = \max_{(l_1,l_2) \in \Lambda_+ \cup \Lambda_-} N_{l_1,l_2}.
\ee
The efficiency of the CT as compared to its full tensor counterparts reads:
\be \label{eq:efficiency}
\frac{N_L^2}{\hat{N}_L}, \quad \text{resp.}\quad \frac{N_L^2}{\hat{N}_L^\text{max}}.
\ee
The matrix operator equations in \eqref{eq:CTsubsystem} can be solved with classical GMRES routines with a matrix-matrix product. The Buffa-Christiansen (BC) function space defines stable pairings and allows for the accurate evaluation of the right-hand side \cite{Hagemann_2019}. Following the same setting as in \eqref{eq:errorSecond}, there exists $L_0(k;\beta)\in\mathbb{N}$
 such that for all $L \geq L_0$ \cite[Theorem 6.3]{JS15_613}:
\be 
\|\Sigma - \hat{\Sigma} _L\|_{X_\beta(\Gamma)^{(2)}}\leq C h_L^s|\log h_L|^\frac{1}{2}\|\Sigma \|_{X_\beta^s(\Gamma)^{(2)}}.
\ee
\section{Numerical Experiments}
\label{sec:Numexp}

\begin{figure*}[!tb]
    \centering
\vspace{-1.2cm}
    \begin{minipage}{0.32\linewidth}
\begin{figure}[H]
\includegraphics[width=0.95\linewidth]{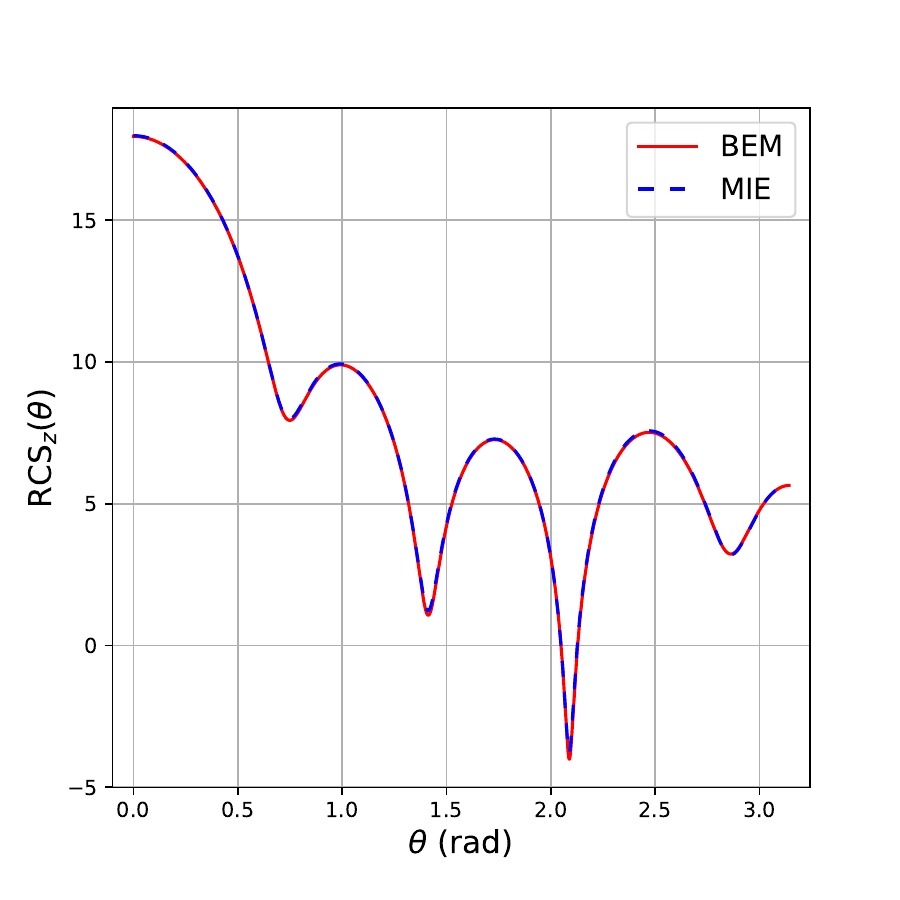}
\vspace{-0.2cm}
\caption*{(a)}
\end{figure}
\vspace{-0.9cm}
\begin{figure}[H]
\includegraphics[width=0.95\linewidth]{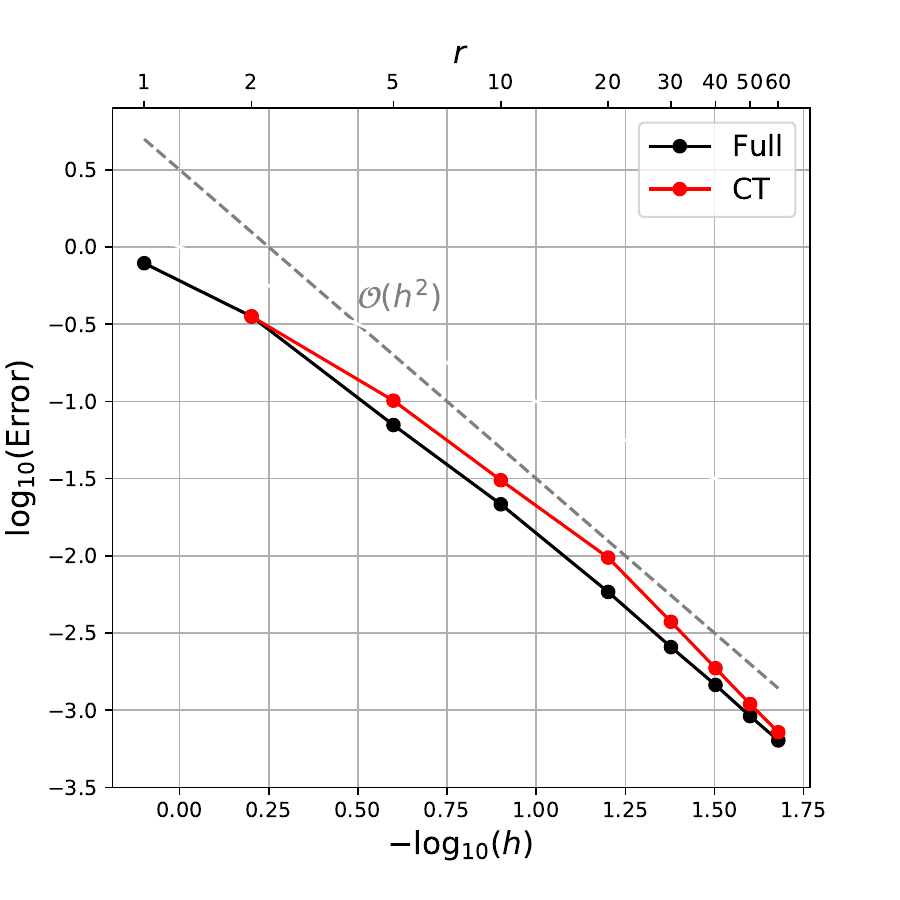}
\vspace{-0.2cm}
\caption*{(d)}
\end{figure}
    \end{minipage}
    \begin{minipage}{0.32\linewidth}
    \begin{figure}[H]
\includegraphics[width=0.95\linewidth]{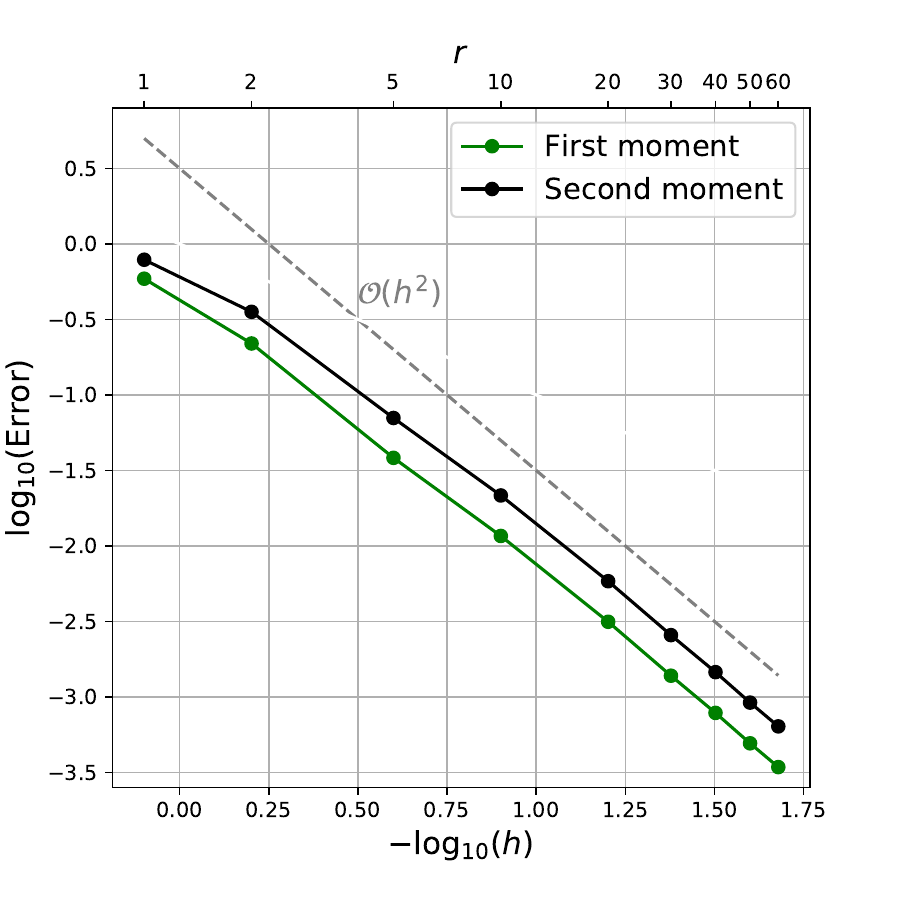}
\vspace{-0.2cm}
\caption*{(b)}
\end{figure}
\vspace{-0.9cm}
\begin{figure}[H]
\includegraphics[width=0.95\linewidth]{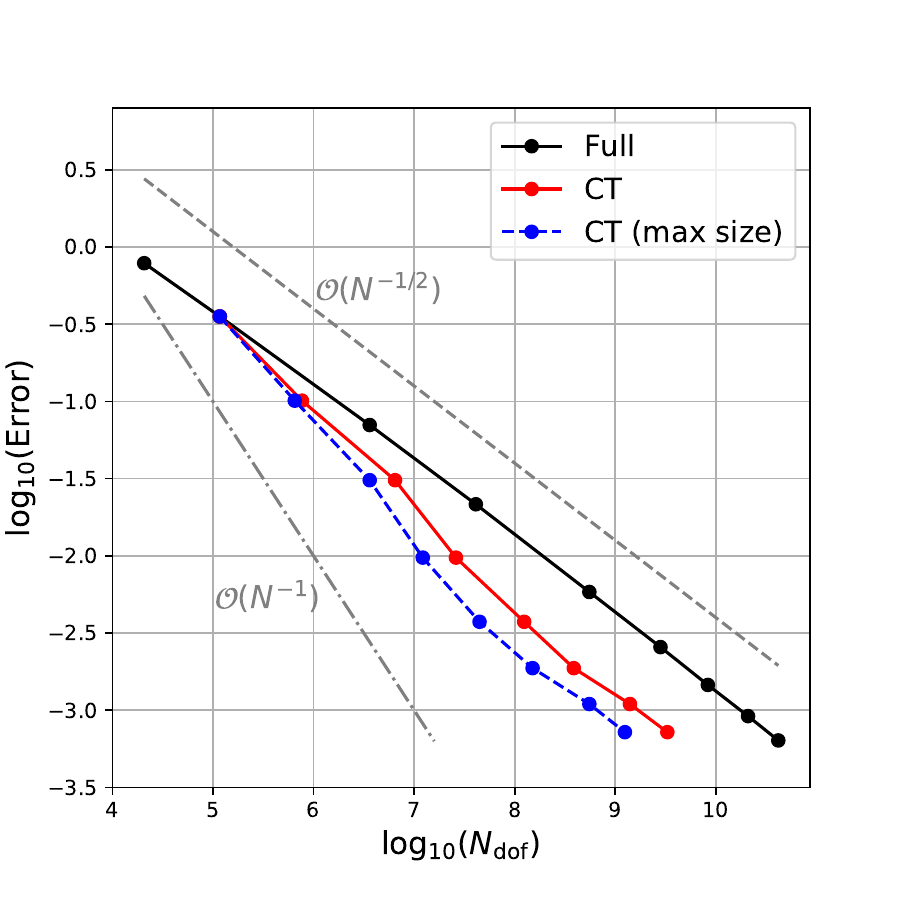}
\vspace{-0.2cm}
\caption*{(e)}
\end{figure}
    \end{minipage}{}
  \begin{minipage}{0.32\linewidth}
    \begin{figure}[H]
\includegraphics[width=0.95\linewidth]{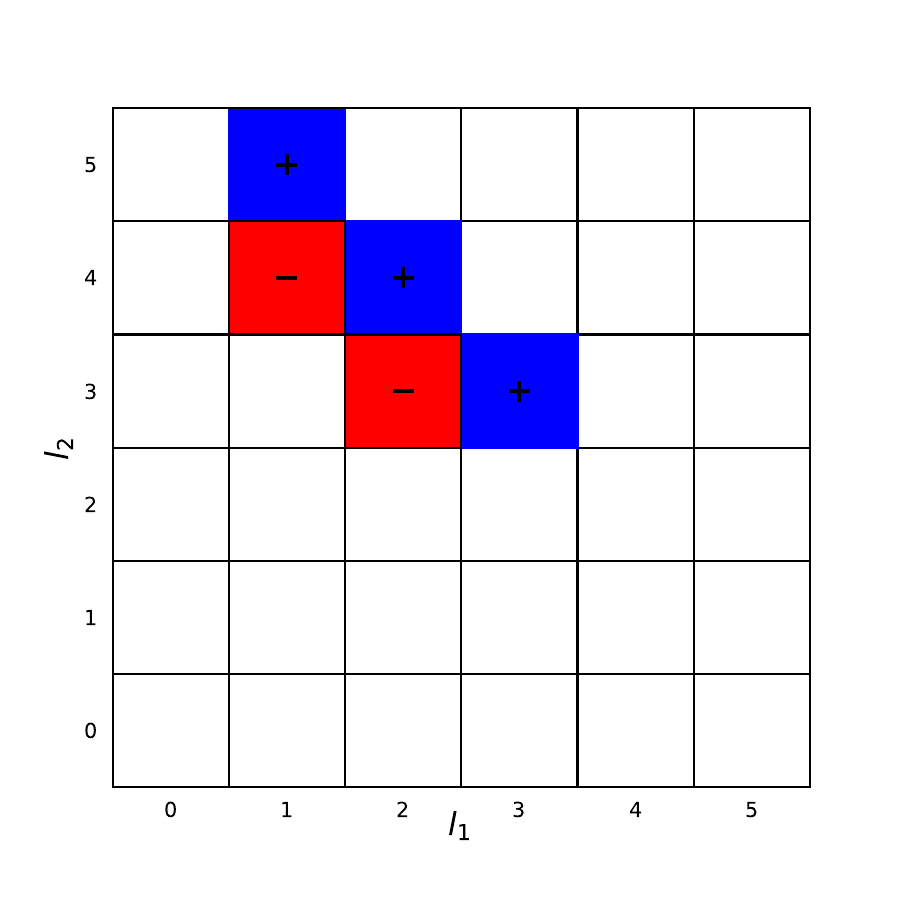}
\vspace{-0.2cm}
\caption*{(c)}
\end{figure}
\vspace{-0.9cm}
\begin{figure}[H]
\includegraphics[width=0.95\linewidth]{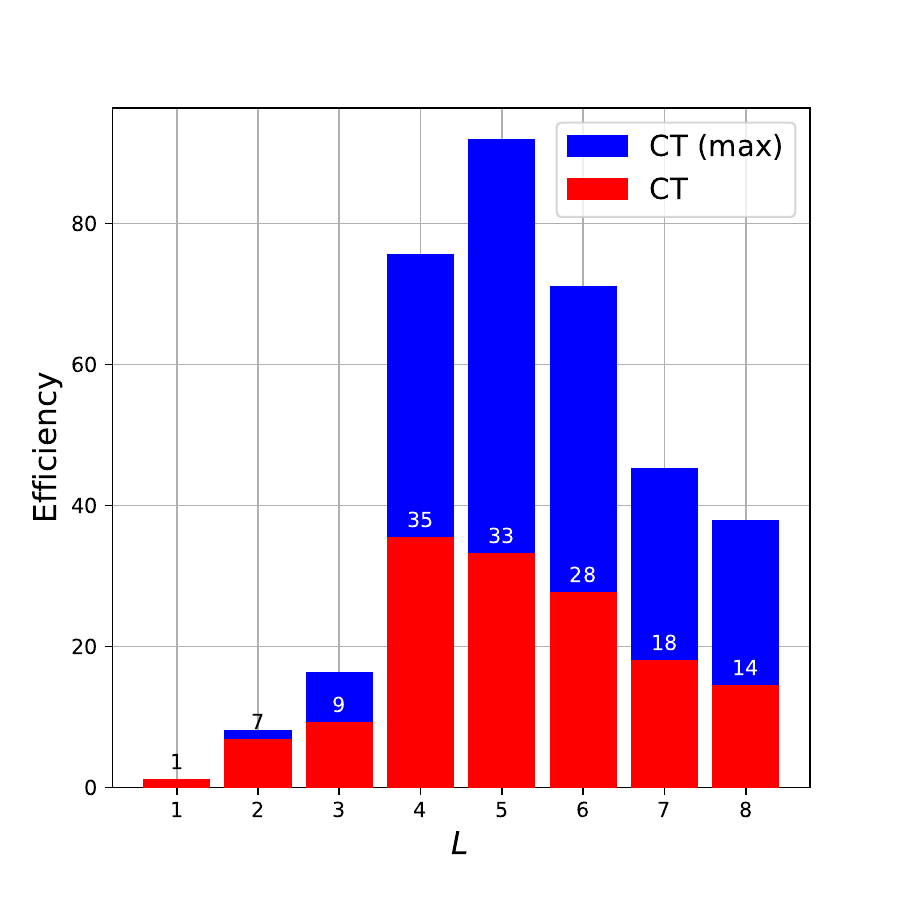}
\vspace{-0.2cm}
\caption*{(f)}
\end{figure}
    \end{minipage}{}
    \caption{Results for unit sphere with known analytic solution and varying problem complexity. (a) $\text{RCS}_z(\theta)$ for $r=10$. (b) Relative error in RCS$_z$ versus $h$: first and second (full) moments. (c) Sub-indices for the CT with hermitian rhs. (d) Relative error in RCS$_z$ versus $h$ for the second moment: Full v/s CT. (e) Relative error in RCS versus $N_\text{dof}$ for the second moment: Full v/s CT. (f) Efficiency of the CT v/s $L$. }
    \label{fig:Sphere}
\end{figure*}

We consider the shape UQ for DE scattering by objects in vacuum as results for PEC are similar. In what follows, we accelerate assembly and matrix-vector products for BEM matrices using $\mH$-mat based on the Adaptive Cross Approximation (ACA) \cite[Subsection 14.2.3]{steinbach_bem_fem}. This has been implemented in the open-source Galerkin boundary element library Bempp 3.3.4 \cite{Bempp}, along with an efficient implementation of Calder\'on preconditioning \cite{IEEE_EJH} using BC function bases. We use default Bempp parameters with ACA tolerance set to $10^{-4}$. Linear systems are solved with GMRES \cite{saad1986gmres}. Tests were executed on a 64 core, 8GB RAM per core, 2xAMD EPYC 7302 server using Python 3.5. 

As stated in \Cref{sec:intro}, our numerical results are fully reproducible via the Bempp-UQ package \cite{FOSB_Helmholtz} following FAIR principles \cite{wilkinson2016fair}. The library is compatible with Bempp Docker image, allowing for further interoperability. It supports all the capabilities of Bempp 3.3.4, allows to conduct shape UQ, compute the SD and solve tensor operator equations.

As in \cite{FOSB_Helmholtz}, we validate our method by tackling the following steps:
\begin{enumerate} 
\item In \Cref{subsec:Numexp_CT} we verify the expected convergence rates for the CT applied to the deterministic DE scattering by the unit sphere with a known analytical solution;
  \item In \Cref{subsec:Numexp_FOA} we check the accuracy of the FOA for the DE scattering by a deterministic kite-shape object;
  \item In \Cref{subsec:Numexp_Kite} we consider the shape UQ for the kite-shape object with a simple random perturbation;
  \item In \Cref{subsec:Complex_case} we consider the Fichera cube and compare our technique to MC method. 
\end{enumerate}
In Bempp-UQ each subsection is within a specific folder, including source codes, datasets and plots here presented. 
\begin{table}[t]
\renewcommand\arraystretch{1.2}
\begin{center}
\footnotesize
\begin{tabular}{|c|c|c|c|c|} \hline   
 \multicolumn{2}{|c|}{Case}&  \Cref{subsec:Numexp_CT} & \Cref{subsec:Numexp_FOA,subsec:Numexp_Kite,subsec:Complex_case}  \\ \hline
\multicolumn{2}{|c|}{$\upmu_r$} & $1.0$ & $1.0$  \\ \hline  
\multicolumn{2}{|c|}{$\upepsilon_r$} & $2.1$  & $1.9$  \\ \hline \hline
\multicolumn{2}{|c|}{$f$} & $143.1$ MHz & $238.6$ MHz \\ \hline
 \multicolumn{2}{|c|}{$\lambda$ (m)} & $2.1$ & $1.26$ \\ \hline 
 \multicolumn{2}{|c|}{$k_0$} & $3.0$ & $5.0$  \\ \hline 
 \multicolumn{2}{|c|}{$k_1$} & $4.3$ & $6.9$ \\ \hline \hline
 \multicolumn{2}{|c|}{$n_\text{angles}$} & $1{,}801$ & $400$ \\ \hline
 \multicolumn{2}{|c|}{tol.~GMRES} &$10^{-8}$ & $10^{-6}$ \\ \hline
\end{tabular}
\end{center} 
\caption{Overview of the relative material parameters, frequency $f$, wavelength $\lambda$ and wavenumbers $k_i$, $i=0,1$ for each case. Next, we detail the number of angles for the RCS evaluation and the relative tolerance for GMRES.}
\label{table:overviewInitial}
\end{table}  
Throughout this section, the objects are illuminated by an incident field:
\be \label{eq:incidentNumexp}
\bE^\text{inc} = \bp e^{\imath k_0 \bd \cdot \bx}.
\ee 
The RCS in \eqref{eq:RCS} is evaluated over $n_\text{angles}$ angles 
$$\theta_i = 2 \pi \frac{i }{n_\text{angles}} , ~ i = 0 ,\cdots, n_\text{angles} - 1.$$
In \Cref{table:overviewInitial}, we show the physical parameters for each case along with $n_\text{angles}$ and GMRES tolerances.

\subsection{Unit-sphere: Convergence analysis}
\label{subsec:Numexp_CT}
Consider as scatterer the unit sphere. We obtain the convergence bounds stated in \Cref{sec:Galerkin} for the Galerkin method and CT. Let the incident field in \eqref{eq:incidentNumexp} be
$$
\bp : =  \begin{pmatrix} 0, 0,  1 \end{pmatrix}^t \quad \text{and}\quad \bd :=  \begin{pmatrix} 1, 0,  0 \end{pmatrix}^t .
$$
We resort to the MIE series \cite{miebookabsorbtion} for the exact solution and check the $z$-component of the far-field at $z=0$. As a tensor BIE, we set a right-hand side $\mM^2[\bff]=  (\bff \otimes \bff)$ with $\bff$ in \eqref{eq:pmchwt}. A straightforward calculation yields that $\mM^2 [\bxi^\text{sc}] = (\bxi^\text{sc} \otimes \bxi^\text{sc})$. 

\begin{table}[t]
\renewcommand\arraystretch{1.2}
\begin{center}
\footnotesize
\resizebox{8.9cm}{!} {
\begin{tabular}{|c|c|c|c|c|c|c|c|c|c|c|} \hline   
$l$ & $0$ & $1$ &$2$ & $3$ & $4$& $5$ & $6$ &$7$ & $8$ \\ \hline 
$r_l$ & $1$ & $2$ &$5$ & $10$ & $20$& $30$ & $40$ &$50$ & $60$ \\ \hline 
 $N_l$ & $144$ & $342$ &$1{,}902$ & $6{,}408$ & $23{,}496$& $53{,}034$ & $91{,}212$ &$144{,}428$ & $204{,}234$ \\ \hline 
\end{tabular}}
\end{center} 
\caption{Unit sphere: Definition of the levels $l = 0, \cdots, 8$. We represent the precision $r_l$ and dofs for the BEM linear system $N_l$ for first moment.}
\label{fig:UnitSphereLevels}
\end{table}  
In \Cref{fig:Sphere}, we represent an exhaustive review of the results.
In \Cref{fig:UnitSphereLevels}, we define the levels $l$ corresponding to precision and degrees of freedom for the first moment. We plot the $\text{RCS}_z$ for $r=10$ in Fig.~\ref{fig:Sphere}(a) for $\theta \in [0 , \pi]$. Next, we verify the $h$-convergence in Fig.~\ref{fig:Sphere}(b). We represent the convergence results for the relative $L^2([0,\pi])$-norm error for the $\text{RCS}_z$ (green). We remark that the error for the first moment is $\mO (h^2)$. The top $x$-axis represents the precision $r$. In the same fashion, we study $ \mM^2[\text{RCS}_z]= \text{RCS}_z^{(2)}$ for he second moment. Its full version yields similar convergence for the relative $L^2([0,\pi])^{(2)}$-norm (black). 

Next, we analyze the accuracy of the CT for the second moment and $L_0 =1$. To begin with, Fig.~\ref{fig:Sphere}(c) represents the sub-blocks for the CT with Hermitian rhs.~for $L=5$ and $L_0=2$. Next, Fig.~\ref{fig:Sphere}(d) compares he convergence for the full tensor (black) and CT (red) for $L_0=1$ and $L = 1,\ldots, 8$. Observe that CT converges similarly to the full tensor case. However, CT uses less dofs. In Fig.~\ref{fig:Sphere}(e) we showcase the accuracy with respect to the total number of dofs. Finally, we show the efficiency \eqref{eq:efficiency} for CT in Fig.~\ref{fig:Sphere}(f) ranging between $6.78$ and $35.40$. Its maximum value is attained for $L=4$ and $L=5$. 

We also study the maximum block size. Here, the efficiency yields further improvement, with values of $75.60$ and $91.81$. This reveals the trade-off between sparsification---increasing $L-L_0$---and the total number of levels, hinting at using $4$ levels.

\subsection{Kite-shaped object: FOA analysis}
\label{subsec:Numexp_FOA}
\begin{figure}[t]
 \centering
\includegraphics[width=\linewidth]{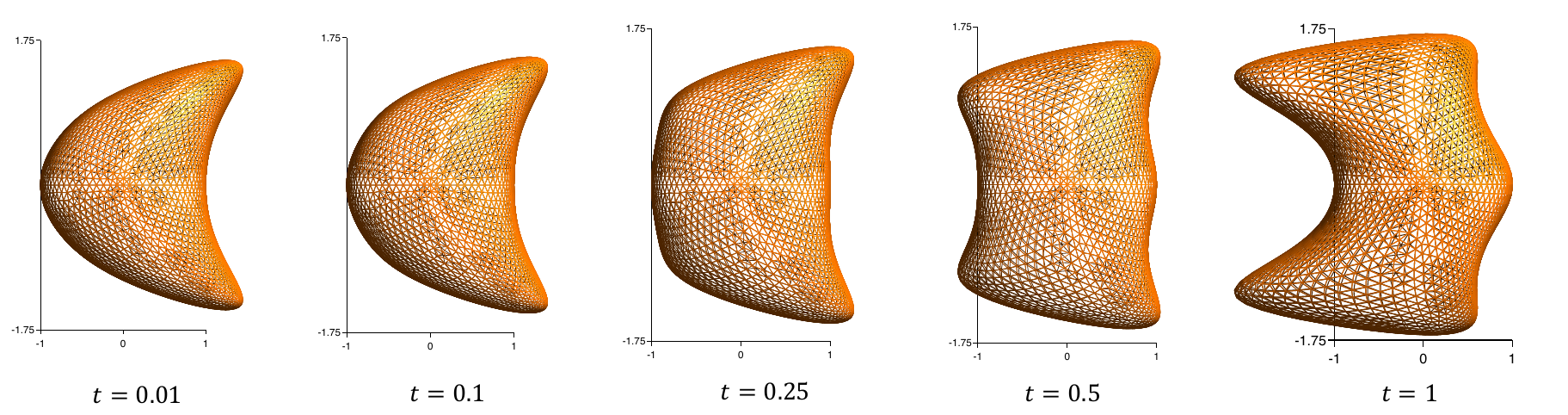}
\vspace{-.2cm}
\caption{Kite-shaped object: Transformed boundaries function to $t$, meshed with $3,249$ vertices.}
\label{figs:transformedBoundariesK}
\end{figure}

\begin{figure}[t]
 \centering
\includegraphics[width=.9\linewidth]{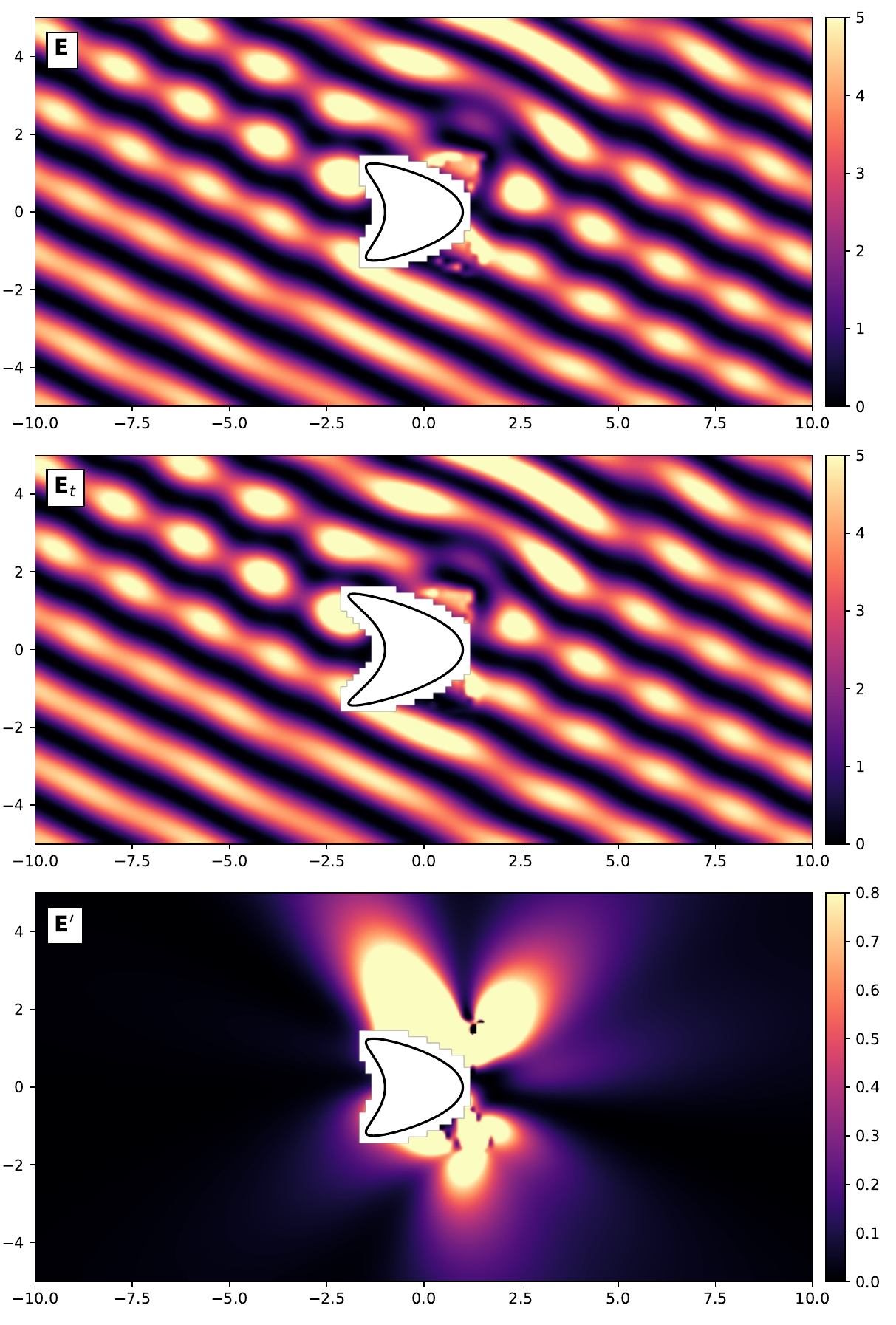}
\caption{Kite-shaped object: Squared electrical density for $\bE$ (top), $\bE_t$ (middle) and $\bE'$ (bottom).}
\label{figs:FOAPlot}
\end{figure}

\begin{figure}[h!tb]
    \centering
\includegraphics[width=0.9\linewidth]{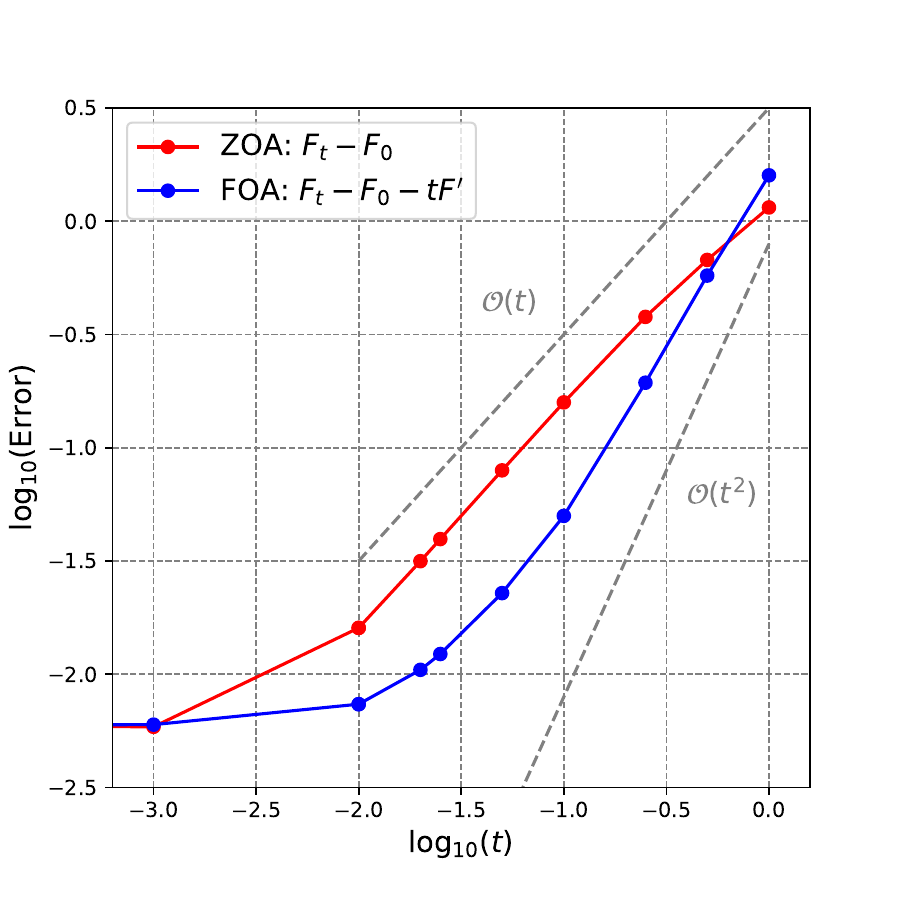}
\caption{FOA v/s ZOA: Results in log-log scale for the kite-shaped object with respect to $t$.}
\label{fig:FOA}
\end{figure}

With the convergence for CT verified, we now check the validity of our FOA. Consider the incident field in \eqref{eq:incidentNumexp} with
$$
\bp : =  \begin{pmatrix} \imath, 2, -1 - \frac{1}{3}\imath \end{pmatrix}^t \quad \text{and}\quad \bd := \frac{1}{\sqrt{14}} \begin{pmatrix} 1, 2, 3 \end{pmatrix}^t.
$$
Inspired by \cite[Section 7.1]{FOSB_Helmholtz}, we introduce a kite-shaped object perturbed deterministically according to $\Gamma_t := \{  \bx + t \overline{\bv}(\bx) ,~ \bx \in \Gamma \}$, with 
\be \label{eq:kitetransfo}
\overline{\bv} (\bx) := \begin{pmatrix} (z^2-1) (\cos(\theta)-1) \\ 0.25 \sin(\theta) (1-z^2) \\ 0\end{pmatrix},
\ee 
for $\theta = \text{atan2}(y,x) \in [0, 2\pi]$. We represent in \Cref{figs:transformedBoundariesK} the family of transformed boundaries considered here, corresponding to $t=\{0.01,0.1,0.25,0.5,1.0\}$. 
To obtain an accurate solution, we set $r=20$ which yields linear systems of $N=9{,}003$ dofs. In \Cref{figs:FOAPlot} we set $t=0.25$ and represent the squared electrical density for $\bE$ (top), $\bE_t$ (middle) and $\bE'$ (bottom). We remark that $\bE$ and $\bE_t$ have similar patterns. The SD displays higher magnitude values close to the scatterer, particularly in the shadow area.
Next, to assess the FOA accuracy, we compare the far-field $\bF_t$ to:
$$
\begin{array}{cc}
\bF, & \textup{zeroth-order approximation (ZOA, in red),}\\
(\bF+ t \bF'), &\textup{FOA (in blue).}\\
\end{array}
$$
Results are portrayed in \Cref{fig:FOA} in log-log scale in relative $L^2([0,2\pi])$-norm. Notice that the error for ZOA scales linearly with $t$, while for FOA it does so quadratically---as $\mO(t^2)$---as expected. Both convergence curves stagnate at $0.60 \%$ for small values of $t$ as the scheme accuracy is reached.

\subsection{Kite-shaped object: Shape UQ}
\label{subsec:Numexp_Kite}
We keep the same parameters as in \Cref{subsec:Numexp_FOA}, but apply a random perturbation as $\Gamma_t (\omega) := \{  \bx + t \bv (\bx,\omega) ,~ \bx \in \Gamma \}$ such that
$$
\bv (\bx, \omega) :=  \mu (\omega)\overline{\bv}(\bx )
$$
with $\overline{\bv}$ in \eqref{eq:kitetransfo} and $\mu \sim \mU[-1,1]$. We set $t=0.05$ and compare results among the following schemes:
\begin{enumerate}
\item MC simulation over $500$ runs;
\item FOA with full tensor equation;
\item FOSB, i.e.~including the CT. 
\end{enumerate}
MC and full tensor systems are solved for $r=20$ whereas CT is performed over three levels corresponding to $r_l \in \{ 5,10,20 \}$. The schemes lead to:
$N_L= 18{,}006$, $N_L^2 = 3.24 \times 10^8$, $\hat{N}_L = 5.38 \times 10^{7}$, $\hat{N}_L^\text{max}=2.54 \times 10^{7}$.

\begin{figure*}[h!tb]
 \centering
\includegraphics[width=.95\linewidth]{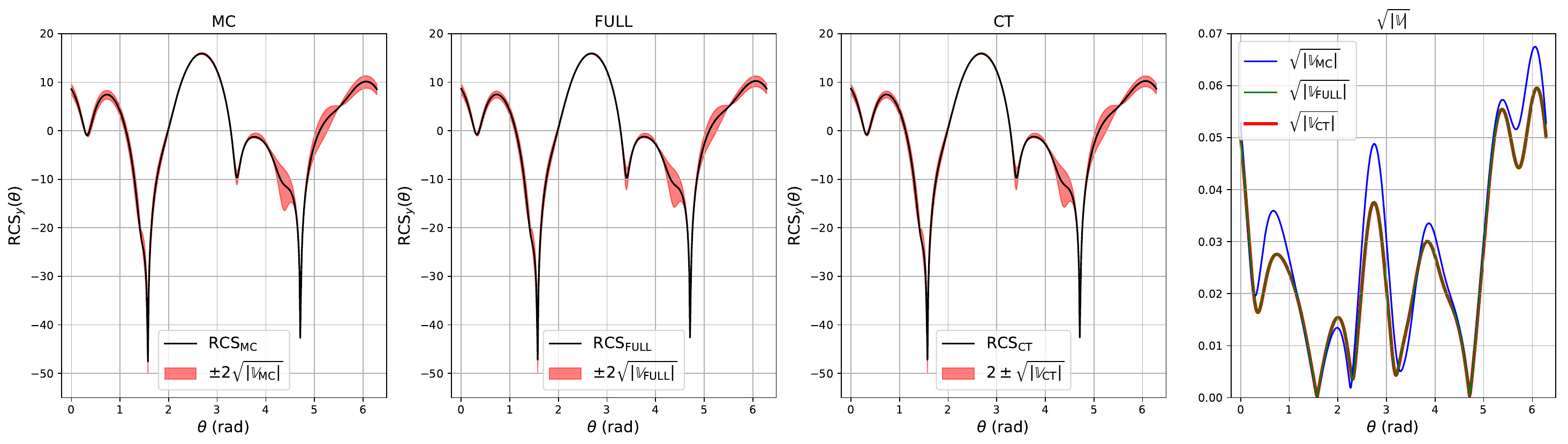}
\caption{Kite-shaped object: Final results for RCS$_y$.}
\label{figs:FOAFinal}
\end{figure*}

In \Cref{figs:FOAFinal}, we show the average and two-standard deviations for RCS$_y$ which reveal that the variance concentrates in the shadow region.
Furthermore, MC and FOA solutions show similar behaviors, differing by a $15.78\%$ and $15.81\%$, respectively. Furthermore, CT yields an approximation to the full version with an efficiency of $6.02$ and $12.77$ with respect to $\hat{N}_L^\text{max}$. The relative error in $L^2([0,2\pi])$-norm between both squared-absolute variances is $8.68 \times 10^{-4}$. On the right-hand figure, we show the squared variance for the three methods, which confirm the accuracy for the FOSB method. 

\subsection{Complex case: Fichera cube}
\label{subsec:Complex_case}
\begin{figure*}[h!tb]
 \centering
\includegraphics[width=.95\linewidth]{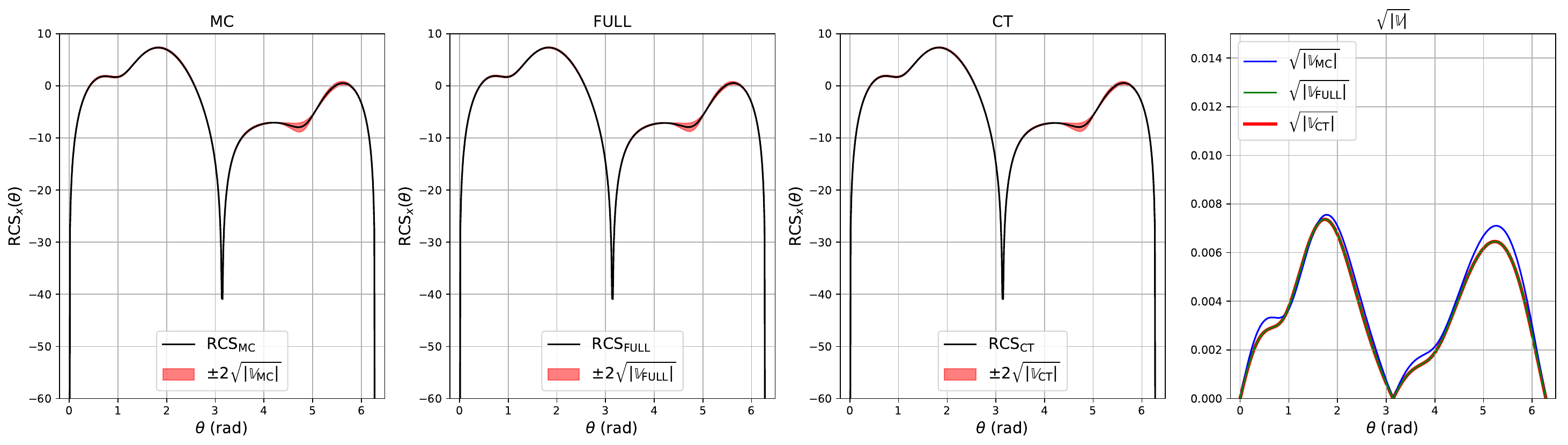}
\includegraphics[width=.95\linewidth]{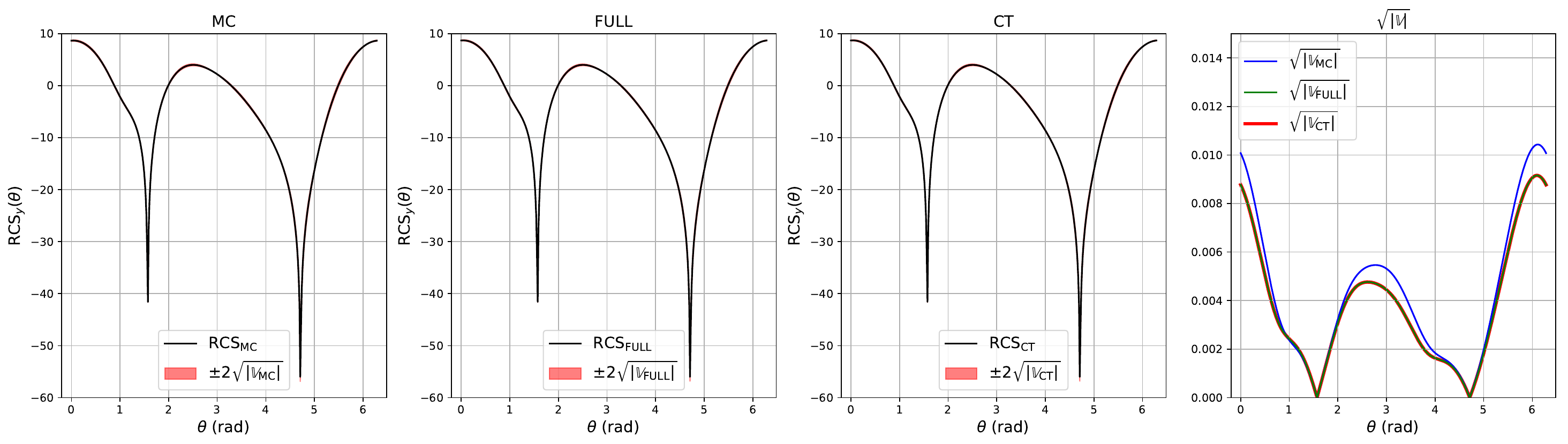}
\includegraphics[width=.95\linewidth]{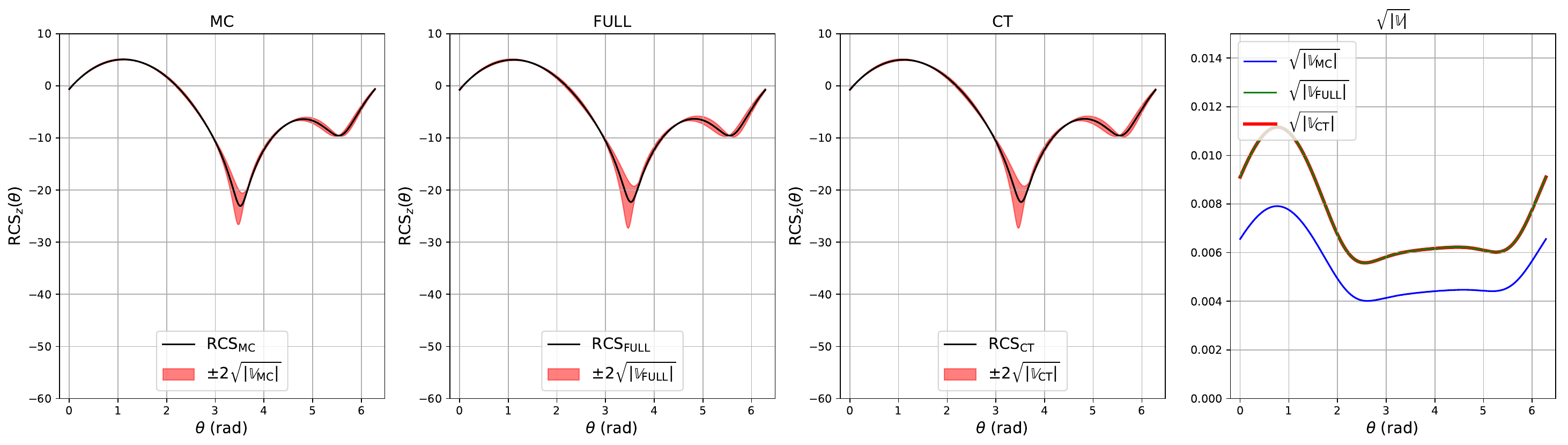}
\caption{Fichera cube: Final results for RCS$_x$ (top) RCS$_y$ (middle) and RCS$_z$ (bottom).}
\label{figs:Fichera}
\end{figure*}

Finally, we consider a non-smooth object, namely the unit Fichera Cube. Similar to \cite{FOSB_Helmholtz}, we perturb the boundary face located at the $z=0.5$-plane along the $z$-axis with unit vector $\hat{\bee}_z$. Given uniformly distributed random variables $\mu_{ij} \in \mU[-1,1]$, $i,j=0,\ldots,5$, the perturbation field is given as:
\be
\label{eq:peturbation}\bv(\bx,\omega):= \sum_{i=0}^5\sum_{j=0}^5 \Upsilon_i(x)\Upsilon_j(y) \mu_{ij} (\omega)\hat{\bee}_z,~ \bx \in \Gamma, ~ z = 0.5
\ee
with $\Upsilon_i$ denoting sine splines of the form $|\sin (q\pi x)|$, $x \in [0,0.5]$, $q\in\{2,4,6\}$ with support of length $0.5 /(q+1)$. Therefore, for $\bx_1$ and $\bx_2$ in $\Gamma$, and $z_1=z_2=0.5$, we have
$$
\mM^{2}[\bv\cdot \bn](\bx_1,\bx_2) = \sum_{i=0}^5\sum_{j=0}^5 \frac{1}{3}\Upsilon_i(x_1)\Upsilon_j(y_1)\Upsilon_i(x_2)\Upsilon_j(y_2).
$$
We set $L_0=0$, $L=2$, yielding three levels for the CT correspond to $r_l \in \{2,5,10 \}$ and $N_l \in\{ 270, 792,3204 \}$ for $l \in \{0,1,2 \}$. We apply MC over $500$ runs and fix $t=0.05$. Provided that the transformation $\bv$ has a Hermitian covariance kernel, CT amounts to computing three sub-block equations (cf.~\cite{FOSB_Helmholtz}). As in \Cref{subsec:Numexp_Kite}, we represent in \Cref{figs:Fichera} the two-band variance for the RCS for MC, full tensor (corresponding to $L=2$) and FOSB solutions. We remark that CT and its full counterpart yield almost identical results. MC and FOA exhibit similar patterns for the $x$ axis (resp.~$y$ axis), with a relative 7.74\% (resp.~12.63\%) square absolute variance difference in $L^2([0,2\pi])$-norm. Along the $z$-axis their magnitudes differ by a 39.31\% (bottom-right). 
The latter hints at the object's regularity resulting in a deterioration of FOA, as it is theoretically set for $C^2$ nominal shapes.
However, the confidence intervals are indistinguishable. 

To conclude, we provide in \Cref{tab:ComprehensiveFichera} comprehensive results pertaining to this particular case. We present execution times per solve and in total and compare memory requirements and solver time along with iteration counts for GMRES. The iterative solver for the tensor equation displays $h$-independent convergence counts. One should  highlight the efficiency of CT: it achieves an efficiency of $5.87$ and a total execution time reduction by a factor of $7.44$. Moreover, this speed-up is even greater as the most intensive sub-block $(2,0)$ is evaluated in $1{,}523$s, whereas the full tensor takes $21{,}944$s, representing a $14.41$ speed-up. Finally, MC takes $70$s approximately and $9$ GMRES iterations for each run with our server processing up to $5$ runs simultaneously. In this setting, it took approximately $7{,}000$s for the MC approximation, which represents a total of $1.75 \times 10^6$ dofs. Thus, FOSB outperforms MC  by $4.60$, as its three sub-systems were processed concurrently.

\begin{table}[t]
\renewcommand\arraystretch{1.2}
\begin{center}
\footnotesize
\begin{tabular}{|c|c|c|c|c|c|c|} \hline   
Method &  \multicolumn{3}{c|}{CT}& Full  \\ \hline
 $(l_0, l_1)$& $(2,0)$  & $(1,1)$ & $(1,0)$ & $(2,2)$   \\ \hline
 dofs&    $8.65\times 10^5$ & $6.27 \times 10^5$ & $2.10 \times 10^5 $   & $1.00\times 10^7$  \\ \hline 
 t$_\text{solve}$ (s)& $1{,}523$& $1{,}057$&  $369$  & $21944$ \\ \hline 
 $n_\text{iter}$ &  $10$ & $9$& $10$& $9$ \\ \hline 
\end{tabular}
\end{center} 
\caption{Fichera cube: FOSB and full FOA performances.}
\label{tab:ComprehensiveFichera}
\end{table}

\section{Conclusions}
\label{sec:conclu}
We carried out shape UQ for full EM wave scattering by PEC and DE objects with small amplitude random perturbations. To reduce computational effort, we applied the FOSB method with provided a FAIR toolbox. Comprehensive numerical experiments demonstrated good applicability and poly-logarithmic computational requirements of the technique. Further work includes: (i) accelerated preconditioning \cite{IEEE_EJH,KLEANTHOUS2022111099,ESCAPILINCHAUSPE2021220} of tensor equations; (ii) low-rank decomposition to the right-hand side for tensor operator equations, e.g.~via pivoted Cholesky factorization \cite{HARBRECHT2012lr}; and (iii) extension to heterogeneous scatterers via local multiple traces formulation \cite{AYALA2022114356} coupled with OSRC preconditioning \cite{fierro2023osrc}. 
\bibliographystyle{IEEEtran}
\bibliography{references}
  

%
\begin{IEEEbiography}[{\includegraphics[width=1in,height=1.25in,clip,keepaspectratio]{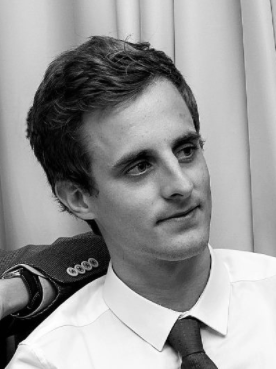}}]{Paul Escapil-Inchauspé} (S'17) obtained his M.Sc.~in Engineering from École Centrale de Nantes, France, in 2015. He completed his B.Sc.~and M.Sc.~in Mathematical and Industrial Engineering, as well as a PhD in Civil Engineering from Pontificia Universidad Católica de Chile (PUC). Currently, he is a Post-Doctoral fellow at the Faculty of Engineering and Sciences, Universidad Adolfo Ibáñez, Santiago, Chile, and works as a Data Science Researcher Engineer at the Data Observatory Foundation. His research interests encompass partial differential equations, uncertainty quantification, physics-informed neural networks, and boundary element methods.
\end{IEEEbiography}
\begin{IEEEbiography}[{\includegraphics[width=1in,height=1.25in,clip,keepaspectratio]{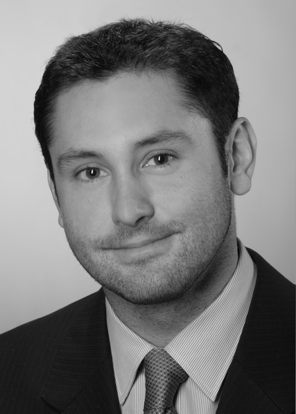}}]{Carlos Jerez-Hanckes} (S'03-M'11) received B.Sc.~and M.Sc.~degrees in electrical and industrial engineering from the Pontificia Universidad Católica de Chile (PUC), Chile, and M.Sc.~and Ph.D.~degrees in applied mathematics from École Polytechnique, France. From 2009 to 2011, he was a Post-doctoral Fellow at ETH Zürich in Switzerland. From 2011 until 2018 he was Associate Professor at the School of Engineering, PUC. Between 2019 and 2022 he was Dean of the Faculty of Engineering and Sciences, Universidad Adolfo Ibañez, Chile, where he is currently Full Professor. His research interests include fast computational methods for acoustic and electromagnetic waves scattering in complex and random media.
Prof.~Jerez-Hanckes is a member of the IEEE UFFC and AP societies.
\end{IEEEbiography}






\end{document}